\documentclass[aps,twocolumn,prd,showpacs,showkeys,preprintnumbers,bibnotes,floatfix,longbibliography,notitlepage,nofootinbib,superscriptaddress,superscriptgaddress,
]{revtex4-2}

\pdfoutput=1
\usepackage{amsmath}
\usepackage{amsfonts}
\usepackage{amssymb}
\usepackage{mathrsfs}
\usepackage{graphicx}
\usepackage{color}
\usepackage{bbding}
\usepackage{tabularx}
\usepackage[usenames,dvipsnames,table]{xcolor}
\usepackage[normalem]{ulem}
\usepackage{makecell}
\usepackage{slashed}

\usepackage[hidelinks]{hyperref}

\hypersetup{colorlinks,citecolor= blue,linkcolor= blue, urlcolor=blue}

\usepackage{multirow}
\usepackage{makecell}
\usepackage{upgreek}
\usepackage[capitalise]{cleveref}

\usepackage{xcolor} 

\begin{document}

\title{Probing axion and flavored new physics with the NA64$\mu$ experiment}

\author{Haotian Li}
\affiliation{Department of Physics, Nanjing University, Nanjing 210093, China}

\author{Zuowei Liu}
\thanks{Corresponding author.\\ zuoweiliu@nju.edu.cn}
\affiliation{Department of Physics, Nanjing University, Nanjing 210093, China}

\author{Ningqiang Song}
\thanks{Corresponding author.\\ songnq@itp.ac.cn}
\affiliation{Institute of Theoretical Physics, Chinese Academy of Sciences, Beijing, 100190, China}


\begin{abstract}

High-energy muon beam dump experiments are powerful probes of new physics models beyond the Standard Model, particularly those involving flavor-dependent interactions. We demonstrate the potential of muon beam dump by placing strong constraints on three new physics models utilizing data from the recent NA64$\mu$ experiment: (1) axions coupling to photons, (2) axions coupling to muons, and (3) a dark sector mediated by a massless $U(1)_{L_\mu-L_\tau}$ gauge boson. The new particles can be identified from the significant missing energy as the invisible channel, or the distinct energy deposition signature as the visible channel.
We find that the current NA64$\mu$ data do not yet probe new parameter region on axion-photon coupling, while excluding new parameter space for the axion-muon coupling $g_{a\mu\mu}\gtrsim4\times10^{-3}$~GeV$^{-1}$ and the axion mass $m_a\lesssim 0.2$~GeV. For the dark sector,  the current data provide stringent constraints that surpass existing ones by nearly one order of magnitude. The data from the 2023 NA64$\mu$ run, once available, will be capable of excluding new axion-photon coupling parameter space and probing the flavor structure of new physics, with more sensitivity advancement expected in near future runs.

\end{abstract}

\maketitle

\section{Introduction}
While the Standard Model (SM) has achieved remarkable success, it fails to explain several key observations, in particular, the existence of dark matter. The extension of SM involves the introduction of new particles or a dark sector. Among them, QCD axion and axion-like particles (ALPs) have drawn increasing attention due to their capability of solving the strong CP problem~\cite{Peccei:1977hh,Peccei:1977ur,Weinberg:1977ma,Wilczek:1977pj} while constituting the cosmological dark matter~\cite{Preskill:1982cy,Abbott:1982af, Dine:1982ah,Duffy:2009ig,Arias:2012az,Semertzidis:2021rxs,Chadha-Day:2021szb,Gu:2021lni,Witten:1984dg,Svrcek:2006yi,Arvanitaki:2009fg}.  We will use axion and ALP interchangeably in this work. 
Axion-photon coupling is constrained from terrestrial haloscopes (e.g.~\cite{ADMX:2009iij,ADMX:2018gho,ADMX:2021nhd}), beam dump and collider experiments (e.g.~\cite{NA64:2020qwq,Belle-II:2020jti,Knapen:2016moh}) and astrophysical observations (e.g.~\cite{Fermi-LAT:2016nkz,Foster:2022fxn,Song:2024rru}). Axions with coupling to fermions such as electrons (e.g.~\cite{PandaX:2017ock, XENON:2022ltv,Capozzi:2020cbu}) and nucleons (e.g.~\cite{Lella:2023bfb,Buschmann:2021juv,Bhusal:2020bvx}) are also widely explored, while axion-muon coupling attracted less attention.

Apart from axions, a natural extension of the SM is a dark sector with a $U(1)$ gauge group. The dark sector is weakly
coupled to the SM by introducing 
kinetic mixing terms \cite{Holdom:1985ag,Foot:1991kb}
between SM gauge bosons and the new $U(1)$ gauge boson, the so-called
``dark photon''~\cite{Batell:2022dpx}. The kinetic mixing may arise through a variety of ways, in particular through loops of heavy particles that are charged under both SM and new $U(1)$ gauge symmetries~\cite{Holdom:1985ag,Cheung:2009qd,Gherghetta:2019coi}. For massless dark photon, the kinetic mixing can be removed through the redefinition of the gauge field, when dark sector particles couple weakly
to SM gauge bosons and appear to be ``millicharged particles'' 
\cite{Feldman:2007wj}. 
Searches for millicharged particles have been conducted in electron and proton beam dumps~\cite{Prinz:1998ua,Golowich:1986tj,Magill:2018tbb,Berlin:2018bsc,Harnik:2019zee}, collider experiments~\cite{Davidson:1991si,Davidson:2000hf,Liu:2018jdi,Liang:2019zkb,Ball:2020dnx}, and through cosmic rays secondaries at neutrino experiments~\cite{Plestid:2020kdm,Kachelriess:2021man,ArguellesDelgado:2021lek,Berlin:2024lwe,Wu:2024iqm,Du:2022hms,Du:2023hsv}.

Despite the extensive
study of axions and the dark sector, the direct probe of flavor-specific new physics has only been available recently with the NA64$\mu$ experiment~\cite{NA64:2024klw,Andreev:2024yft}, where a muon beam is dumped onto the target to search for missing energy. New runs have been made in 2023 to increase the number of muons. Although the electroweak interaction in the SM is flavor independent, flavor non-universality is generally expected in supersymmetry and flavored Higgs models~\cite{Altmannshofer:2012ar,Altmannshofer:2015esa,Evans:2015swa,Bauer:2015fxa,Jiang:2024cqj,Ariga:2023fjg,Galon:2019owl,Abdughani:2021oit,Lu:2023ryd}. In particular, lepton flavor non-universality has been studied in the context of non-standard neutrino interactions~\cite{Coloma:2020gfv,Gninenko:2020xys} and flavor-violating interactions~\cite{Bauer:2019gfk,Bauer:2020itv,Bauer:2021mvw,Batell:2024cdl,Gninenko:2018num,Gninenko:2022ttd,Radics:2023tkn}. A muon beam dump is hence one of the superior ways of probing flavor-dependent new physics, and it is timely to investigate the current status and prospects of new physics with NA64$\mu$.

\begin{figure*}
    \centering
    \includegraphics[width=0.75\linewidth]{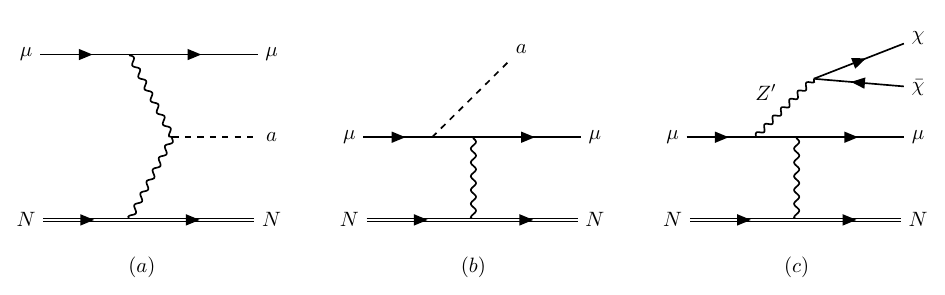}
    \caption{Feynman diagrams for the production of dark particles in the NA64$\mu$ experiment: (a) axion production through photon-photon fusion, (b) axion production through muon bremsstrahlung, and (c) a dark sector where the dark particle $\chi$ is produced through a massless mediator $Z'$.}
    \label{fig:FeyD-production}
\end{figure*}

To demonstrate the full potential of a beam dump in exploring axion and flavor-specific physics, we study three scenarios illustrated in Fig.~\ref{fig:FeyD-production}: 1) axion coupling to photons, 2) axion coupling to muons, and 3) a dark sector interacting with muon through massless $U(1)_{L_\mu-L_\tau}$ gauge boson, which appears to be millicharged in other beam dump experiments. We derive the corresponding constraints on dark particles with mass below GeV by investigating the missing energy events or the distinct visible signal at the NA64$\mu$ experiment, which cover a large parameter space that 
was unexplored before. The sensitivity will be further improved in ongoing searches with higher muon luminosity.

\section{The new physics signals at NA64$\mu$}
The NA64$\mu$ experiment is conducted at CERN using the M2 beam line. The muon beam (with the momentum of 160$\pm$3~GeV)
from the proton dump is selected and collimated to be incident 
on the target, which is the electromagnetic calorimeter (ECAL) 
consisting of lead and plastic scintillator layers \cite{NA64:2024klw,Andreev:2024yft}. 
After exiting ECAL, the muon traverses the veto counter, hadronic calorimeter (VHCAL), muon trackers and finally two large hadronic calorimeters (HCALs). We use the data sample taken at NA64$\mu$ in May 2022~\cite{NA64:2024klw}. The data are recorded as the calorimeter-deposited energy ($E_{\rm CAL}$) and the outgoing muon momentum ($p_{\rm out}$), where different categories of events lie in different regions of the $E_{\rm CAL}-p_{\rm out}$ plane. 

We consider two types of signal signatures: visible and invisible. The visible signal region is defined by the energy conservation condition $E_{\rm CAL}+p_{\rm out}\simeq 160~\rm GeV$, corresponding to events where the new particle escapes the ECAL but decays before leaving the last calorimeter, similar to the visible channel in NA64$e$~\cite{Dusaev:2020gxi,NA64:2020qwq}. The invisible events take place when the new particle decays beyond the last calorimeter, or the new particle does not deposit energy in the calorimeters.
We choose the invisible signal region $p_{\rm out}<120~\mathrm{GeV}$ and $E_{\mathrm{CAL}}<12~\mathrm{GeV}$. No events are observed~\cite{NA64:2024klw} in this region and the expected number of SM background events is also much less than 1, which can be safely neglected.

As depicted in Fig.~\ref{fig:FeyD-production}, we consider the production of dark particles through the photon-photon fusion and bremsstrahlung-like processes at NA64$\mu$. The expected number of beyond-Standard Model (BSM) events is generally computed by
\begin{equation}
    N_{\mathrm{signal}}=N_{\mathrm{MOT}} n_{\mathrm{Pb}}
    L_T\int d\sigma_{\mu N\to\mu NX}\kappa P_{\rm inv(v)}\,,
    \label{eq:Nsignal}
\end{equation}
where $N_{\mathrm{MOT}}=2\times10^{10}$ is the number of muons on target (MOT), 
$n_{\mathrm{Pb}}=3.3\times10^{22}\ \mathrm{cm}^{-3}$ and $L_T = 20\ \mathrm{cm}$ are the number density and thickness of the lead target~\cite{NA64:2024klw}. 
Here conservatively we have ignored the contributions from the plastic scintillator layers of the ECAL.
$\sigma$ is the model-specific cross section, and $\kappa$ the signal efficiency. 
The efficiency depends on the effective mass of the bremsstrahlung particle, which we infer from the results in~\cite{NA64:2024klw} and agrees with the shape in~\cite{Andreev:2024yft}. If the particle is off-shell, the effective mass is the momentum transfer to the particle $\sqrt{Q^2}$. 
If the dark particle is unstable, it may decay to final states that are invisible or visible to the calorimeters described by the probability $P_{\rm inv(v)}$. 

\section{Axion-photon coupling}
We first consider axion with photon coupling, which is common to QCD axions. The corresponding Lagrangian is
\begin{equation}
     \mathcal{L}\supset \frac{1}{2}\partial_\mu a\partial^\mu a-\frac{1}{2}m_a^2a^2-\frac{1}{4}g_{a\gamma\gamma}aF_{\mu\nu}\tilde{F}^{\mu\nu}\,,
\end{equation}
where $F_{\mu\nu}$ denotes the strength tensor of the photon field and $\tilde{F}_{\mu\nu}$ is its dual $\tilde{F}_{\mu\nu}=\frac{1}{2}\epsilon_{\mu\nu\rho\sigma}F^{\rho\sigma}$. $m_a$ and $g_{a\gamma\gamma}$ are axion mass and its coupling to photon.

As described in Fig.~\ref{fig:FeyD-production}, axions are 
mainly produced through the photon-photon fusion. We use the Weizs\"acker-Williams (WW) approximation~\cite{vonWeizsacker:1934nji,williams1935correlation,Kim:1973he,Tsai:1973py}, which simplifies the phase space integration of the $2\to3$ process by treating the virtual photon mediator that is attached to $N$
as a real photon, reducing it to that of a $2\to2$ process. The approximation works well in the relativistic and collinear limit, particularly in the beam dump experiment~\cite{Liu:2017htz} and for the purpose of this work~\cite{Sieber:2023nkq}.  Under the WW approximation, the differential cross section of the axion production process can be written as \cite{Liu:2017htz,Liu:2023bby}
\begin{equation}
    \frac{\mathrm{d}\sigma}{\mathrm{d}x}=\frac{\alpha}{8\pi^2}\sqrt{E_a^2-m^2_a}E_\mu(1-x)\int\mathrm{d}\cos\theta\frac{\zeta}{\tilde{u}^2}\mathcal{A}\,,
    \label{eq:diffxs}
\end{equation}
where $x=E_a/E_\mu$, $\theta$ is the angle between the dark state particle and the beam. 
Numerical study suggests that the integral could take the $\theta$ range from 0 to 0.1~\cite{Sieber:2023nkq}.
$\tilde{u}$ is the modified Mandelstam variable 
$\tilde{u}= (p_{\mu,\rm in}-p_a)^2-m_\mu^2 
=-xE^2_\mu\theta^2-m^2_a\frac{1-x}{x}-m^2_\mu x$, where $p_{\mu,\rm in}$ and $p_a$ are the four momenta of the incoming muon and outgoing axion. $\zeta$ is the effective photon flux defined by
\begin{equation}\label{eq:chi}
\zeta=\int^{t_{\mathrm{max}}}_{t_{\mathrm{min}}}\mathrm{d}t\frac{t-t_{\mathrm{min}}}{t^2}F^2(t)\,,
\end{equation}
where the squared momentum transfer to the nucleus $t$ varies between $t_{\mathrm{min}}\approx\tilde{u}^2/(4E^2_\mu(1-x)^2)$ and $t_{\mathrm{max}}\approx m^2_a+m^2_\mu$~\cite{Kirpichnikov:2021jev}. 
$F(t)$ is the elastic form factor of the nucleus~\cite{Bjorken:2009mm}
\begin{equation}
    F(t)\simeq Z\left(\frac{b^2t}{1+b^2t}\right)\left(\frac{1}{1+t/d}\right)\,,
\end{equation}
with $b=111Z^{-1/3}/m_e$ and $d=0.164~\mathrm{GeV}^2A^{-2/3}$. For lead target the atomic number $Z=82$ and the mass number $A=207.2$. $\mathcal{A} $ is the spin summed and averaged matrix element of the $2\to2$ process in the limit $t=t_{\mathrm{min}}$. 
For axion production through photon-photon fusion this is 
\begin{equation}
    \mathcal{A}=-e^2g^2_{a\gamma\gamma}\tilde{u}^2\frac{\tilde{u}x(2-x)+2m^2_\mu x^2+m^2_a(1-x)(2-x)}{(m^2_a(1-x)+x\tilde{u})^2}\,.
\end{equation}

We then compute the production of axion using Eq.~\eqref{eq:Nsignal}. As axion is generated on-shell, $Q^2=m_a^2$. It is worth mentioning that the energy loss of muon is negligible in the target~\cite{Chen:2017awl}, and we keep the incoming muon energy as a constant. The outgoing axion then leaves the target calorimeter and decay to photons. We focus on axions with mass $m_a\lesssim 0.1$~GeV. 
As axion with gauge boson coupling decays dominantly to photons~\cite{Alonso-Alvarez:2018irt}, we neglect other decay channels.
Although the axion production probability is uniform in the target, production location does affect the propagation distance before axion decay. We take this into consideration by summing up the decay probability for axion to be produced in different layers in the target, as detailed in Appendix~\ref{sec:appendix-decay}.

We derive the current constraints using the visible channel in the signal box $50~\mathrm{GeV}<p_{\rm out}<80~\mathrm{GeV}$ where four events were observed~\cite{NA64:2024klw}, corresponding to the upper limit of 7.8 events at 90\% CL. The background could be substantially mitigated in the future when the energy deposition at each calorimeter can be accessed individually in addition to the total energy deposition. 
Most SM events will deposit energy either promptly in the ECAL, or in the HCALs in the form of a muon. In contrast, axion decay will result in large energy deposition in the HCALs that is incompatible with the muon energy measured by the tracking detectors before and after the HCALs. One exception is the hadronic process secondaries produced in the SM, which may mimic the behavior of axion decay. However, such background can also be largely vetoed by applying appropriate event selection cuts, similar to the search for axion decay in NA64e~\cite{NA64:2020qwq}.
We therefore assume background-free in the projection of the visible channel. For the invisible channel, we require the axion to decay after the end of the last HCAL.

\begin{figure}
    \centering
    \includegraphics[width=\linewidth]{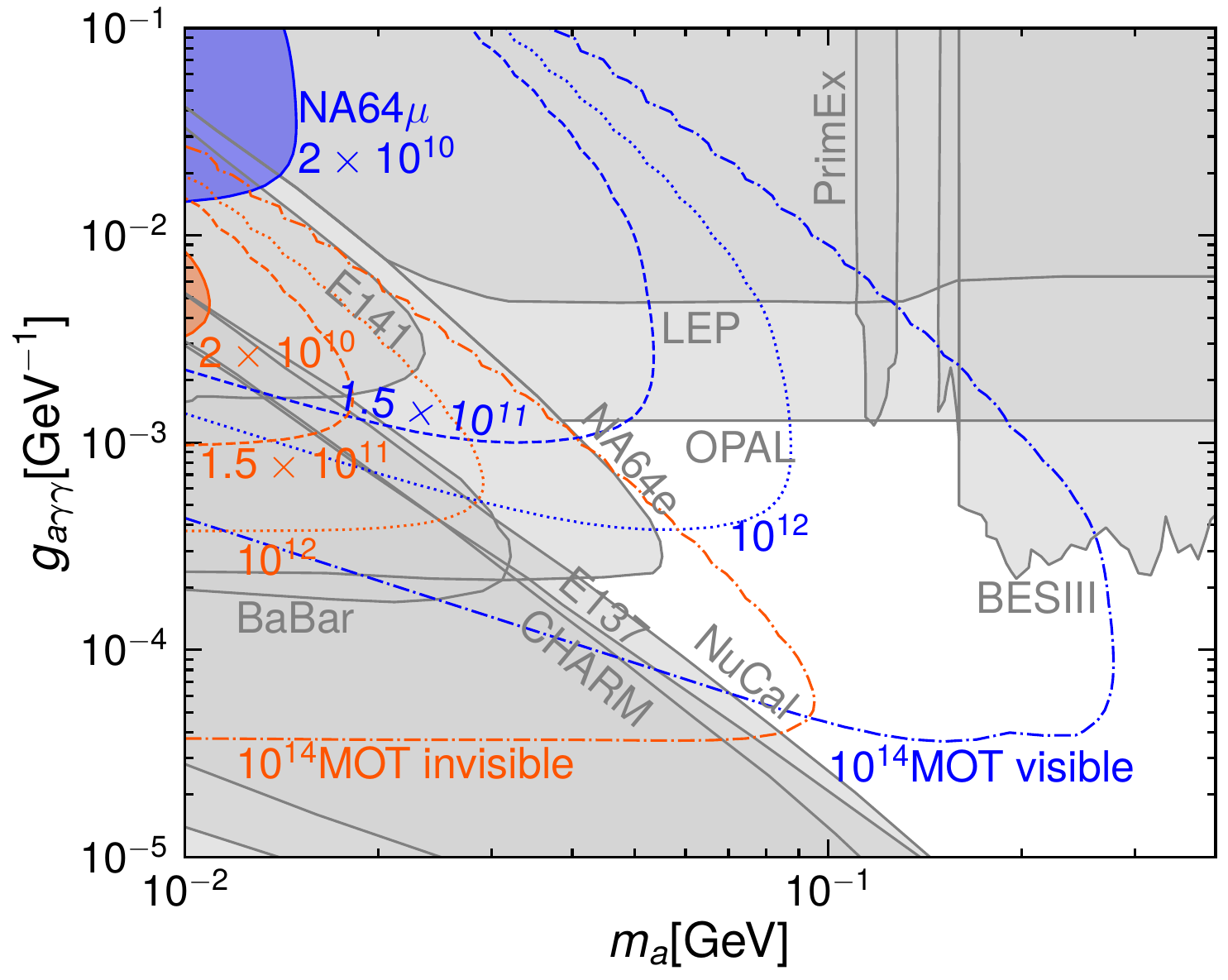}
    \caption{90\% C.L. limits on the axion-photon coupling with the current NA64$\mu$ data (blue shaded region for the visible channel and orange shaded region for the invisible channel), and the projected sensitivity with $1.5\times10^{11}$, $10^{12}$ and $10^{14}$ muon on target (dashed blue lines for the visible channel and orange lines for invisible). Existing constraints from collider and beam dump experiments are shown as grey shaded regions, including LEP~\cite{Jaeckel:2015jla}, OPAL~\cite{Knapen:2016moh}, PrimEx~\cite{Aloni:2019ruo}, BESIII~\cite{BESIII:2022rzz,BESIII:2024hdv}, E141 \cite{Dobrich:2017gcm}, NA64 electron beam dump~\cite{NA64:2020qwq}, BaBar \cite{Dolan:2017osp}, NuCal \cite{Blumlein:1990ay}, E137 \cite{Bjorken:1988as} and CHARM \cite{CHARM:1985anb}. }
    
    \label{fig:axion-photon}
\end{figure}

We show the constraints and sensitivities in $g_{a\gamma\gamma}$ in Fig.~\ref{fig:axion-photon}. The current limits are inferior to the existing ones. However, vast improvements are expected in the future with more MOTs, as discussed in Sec.~\ref{sec:conclusions}.

\section{Axion-muon coupling}
We also consider axion (ALP) coupling to muons with the Lagrangian
\begin{equation}
     \mathcal{L}\supset \frac{1}{2}\partial_\sigma a\partial^\sigma a-\frac{1}{2}m^2_aa^2-\mathrm{i}g_{a\mu\mu}(2m_\mu) \bar{\mu}\gamma_5\mu a\,,
\end{equation}
where $g_{a\mu\mu}$ is the axion-muon coupling. 
Under WW approximation, the differential cross section for axion production in muon bremsstrahlung is also given by Eq.~\eqref{eq:diffxs}, with the matrix element
\begin{equation}
  \mathcal{A}=4e^2g^2_{a\mu\mu}m^2_\mu\left[\frac{x^2}{1-x}+2m^2_a\frac{\tilde{u}x+m^2_a(1-x)+m_\mu^2x^2}{\tilde{u}^2}\right]\,.  
\end{equation}
 
\begin{figure}
    \centering
    \includegraphics[width=\linewidth]{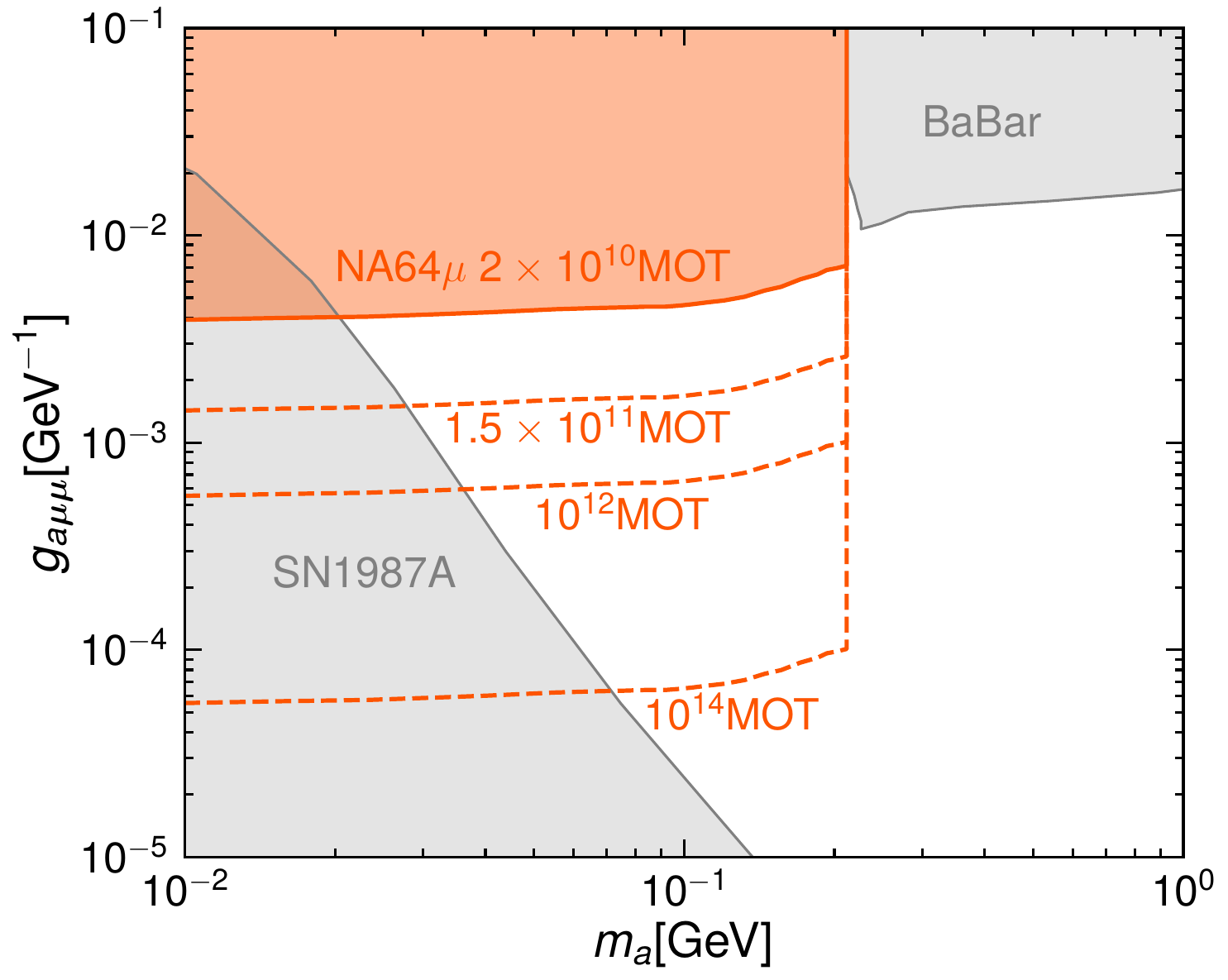}
    \caption{Same as Fig.~\ref{fig:axion-photon} but for axion-muon coupling using the invisible channel.
    Existing constraints are shown as grey shaded regions, including BaBar \cite{BaBar:2016sci} and SN1987A \cite{Croon:2020lrf,Bollig:2020xdr,Caputo:2021rux,Caputo:2022rca}. 
    }
    \label{fig:axion-muon}
\end{figure}

The constraints on axion-muon coupling are shown in Fig.~\ref{fig:axion-muon} using the invisible channel.  
For $m_a<2m_\mu$, axion decay to muons is kinematically forbidden. In the absence of direct coupling to other SM particles, the decay to other particles is loop suppressed. Consequently the produced axions are long-lived which will not deposit energy in the detector, consistent with the signal region in this work. Heavier axion dominantly decays to two muons. 
Conservatively, we require axion to decay beyond the last HCAL calorimeter to satisfy the trigger condition. The resultant sensitivity is not competitive and outside the scope of the figure. A similar pseudoscalar model is considered in~\cite{Sieber:2023nkq} where the sensitivity is projected assuming the pesudosclalar decays to dark sector particles without the inclusion of the efficiency. The sensitivity for the scalar-type coupling is projected in~\cite{Chen:2017awl,Kahn:2018cqs,Gninenko2018NA64mu}.

The NA64$\mu$ experiment with its current exposure could exclude large parameter space for $m_a\lesssim 0.2$~GeV.
Lower axion-muon coupling is constrained by the cooling of SN1987A~\cite{Croon:2020lrf}. At higher mass it is constrained by four-muon final states at BaBar~\cite{BaBar:2016sci}.  

\section{Flavor-dependent dark sector}
Finally we consider the coupling between muon and the dark sector through a massless mediator. This is analogous to millicharged particles where the dark sector particles are lightly coupled to photon and can be produced through the bremsstrahlung of charged particles. We now introduce a flavor-nonuniversal mediator so that the dark particle $\chi$ is preferably produced in muon scattering, while the coupling of $\chi$ to other particles is loop suppressed. This could be realized through a scalar or vector mediator. For concreteness, we consider the vector mediator to be a $L_\mu-L_\tau$ gauge boson, the interaction Lagrangian is given by
\begin{equation}
 \mathcal{L}\supset g_{Z'}J_{\mu-\tau}^\sigma Z'_{\sigma}+g_\chi\bar{\chi}\slashed{Z'}\chi
\end{equation}
with
\begin{equation}
J_{\mu-\tau}^\sigma=\bar{\mu}\gamma^\sigma\mu+\bar{\nu}_{\mu L}\gamma^\sigma\nu_{\mu L}-\bar{\tau}\gamma^\sigma\tau-\bar{\nu}_{\tau L}\gamma^\sigma\nu_{\tau L}
\end{equation}
where the massless gauge boson $Z'$ interacts with the dark fermion $\chi$ via coupling $g_\chi$, and $g_{Z'}$ is the coupling of $Z'$ to the second and third generation leptons. 

As in Fig.~\ref{fig:FeyD-production}, $\chi$ is produced in the $2\to4$ process $\mu N\to\mu N\chi\bar{\chi}$. The differential cross section of this  process can be written as~\cite{Gninenko:2018ter}
\begin{equation}
\begin{split}
    \mathrm{d}\sigma_{\mu N\to\mu N\chi\bar{\chi}}=&\mathrm{d}\sigma_{\mu N\to\mu NZ'}\\
    &\times\frac{g^2_\chi}{12\pi^2}\frac{\mathrm{d}Q^2}{Q^2}\sqrt{1-\frac{4m^2_\chi}{Q^2}}\left(1+\frac{2m^2_\chi}{Q^2}\right)\,,
\end{split}
\end{equation}
where $\mathrm{d}\sigma(\mu N\to\mu NZ')$ is the differential cross section with a virtual $Z'$ in the final state, and $Q^2$ the squared 4-momentum of the virtual $Z'$.
Likewise, $\mathrm{d}\sigma(\mu N\to\mu NZ')$ is given by Eq.~\eqref{eq:diffxs} with $E_a$ replaced by $E_{Z'}$ and $m^2_a$ replaced by $Q^2$. The matrix element is now
\begin{equation}
\begin{split}
     \mathcal{A}=&4e^2g^2_{Z'}\Big[\frac{x^2-2x+2}{2(1-x)}+\frac{Q^2+2m_\mu^2}{\tilde{u}}\\
     &+\frac{2m_\mu^4x^2+Q^4(1-x)+m_\mu^2Q^2(x^2-2x+2)}{\tilde{u}^2}\Big], 
\end{split}
\end{equation}
where $\tilde u =-xE^2_\mu\theta^2-Q^2\frac{1-x}{x}-m^2_\mu x$.
Since the dark fermion is invisible to the calorimeters, $P_{\rm inv}=1$ for the whole mass range we consider.

\begin{figure}
    \centering
    \includegraphics[width=\linewidth]{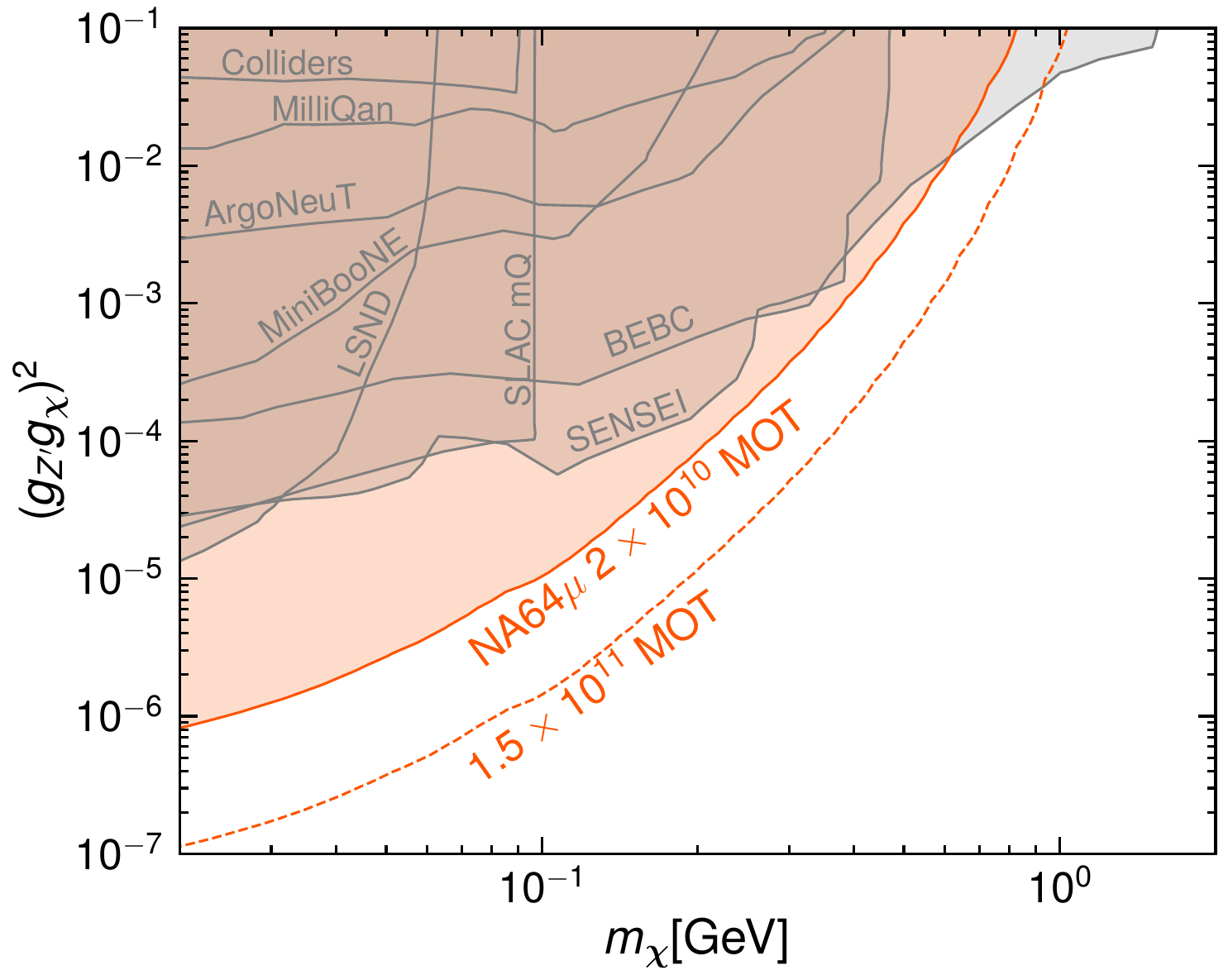}
    \caption{
    Same as Fig.~\ref{fig:axion-photon} but for a flavor-dependent dark sector using the invisible channel.
    We recast the existing limits of millicharged particles in the grey shade regions, including Colliders \cite{Davidson:2000hf}, MilliQan \cite{Ball:2020dnx}, ArgoNeuT \cite{ArgoNeuT:2019ckq}, MiniBooNE \cite{Magill:2018tbb}, LSND \cite{Magill:2018tbb}, SLAC mQ \cite{Prinz:1998ua}, BEBC \cite{Marocco:2020dqu} and SENSEI \cite{SENSEI:2023gie}.}
    \label{fig:Zchi}
\end{figure}

The constraint obtained from the current NA64$\mu$ data using the invisible channel is shown in Fig.~\ref{fig:Zchi}, compared with existing experiments where $\chi$ is produced in electron or proton collisions through electromagnetic or weak interaction. The SM photon couples to $Z'$ through the muon and tau loop, with the effective kinetic mixing~\cite{Araki:2017wyg} 
\begin{equation}
    \Pi(q^2)=\frac{eg_{Z'}}{2\pi^2}\int^1_0\mathrm{d}x (1-x)\ln\frac{m^2_\tau-x(1-x)q^2}{m^2_\mu-x(1-x)q^2}\,.
\end{equation}
By redefining the field to remove the kinetic mixing, $\chi$ couples to photon direct with the interaction $ \mathcal{L}_{\mathrm{int}}\supset g_\chi\Pi(q^2)\bar{\chi}\slashed{A}\chi$. Therefore, the dark fermion $\chi$ can be seen as millicharged particle in these production processes. In the original millicharge model, if the dark particle carries the charge $\epsilon e$, it is equivalent to the flavor-dependent model with the coupling $g_{Z'}g_\chi=\epsilon eg_{Z'}/\Pi(q^2)$. For a massless mediator, the production of $\chi$ is dominated by the minimum possible momentum transfer allowed by kinematics. We then choose $q^2=4m_\chi^2$ when recasting the constraint from $\epsilon e$ to $g_{Z'}g_\chi$.

Although existing experiments have placed strong constraints on the dark particle millicharge, the sensitivity is lost severely by introducing the loop to interact with muon in the flavor-dependent model. The current NA64$\mu$ data leads to more stringent constraints on $(g_{Z'}g_\chi)^2$ than existing ones by up to more than an order of magnitude in the mass range $m_\chi\lesssim 1$~GeV. 
We do not show the bounds from SN1987A, which constrain the parameter region  
$(g_{Z'}g_\chi)^2 \lesssim 4\times10^{-9}$~\cite{Chang:2018rso,Li:2024pcp} and 
therefore lie outside the range plotted in Fig.~\ref{fig:Zchi}.
Stringent constraints were also placed using compact binary systems~\cite{KumarPoddar:2019ceq}; however, 
for 
$g_{Z'}\gtrsim 10^{-18}$~\cite{Liu:2025zuz}, the additional force introduced by $Z'$ will strongly suppress the muon abundance, rendering such constraints unreliable.

\section{Conclusions and prospects}
\label{sec:conclusions} 
The current NA64$\mu$ experiment with its pilot run has already achieved superior constraints on BSM physics, in particular flavor-specific physics than previous experiments. It is followed by a new run at CERN with $1.5\times 10^{11}$ MOT utilizing better magnet spectrometer, detection and trigger systems in 2023. NA64$\mu$ plans to accumulate $10^{12}$ MOT before the long shutdown to explore more parameter space~\cite{NA642023report}, and eventually reach the goal of $10^{14}$ MOT~\cite{Andreev:2024yft,Gninenko:2025aek}. We note that the final MOT are subject to change depending on the deployment and status of the experiment in the future.

We present the projected sensitivity for the 2023 run and for $10^{12}$ and $10^{14}$ MOT at NA64$\mu$ with a similar experimental setup in Fig.~\ref{fig:axion-photon} through Fig.~\ref{fig:Zchi}. The axion-photon coupling does not scale linearly as $N_{\rm MOT}^{-1/2}$ as the coupling also affects the lifetime of axion. More muons in the beam allow probing axion mass up to about 0.08~GeV at $g_{a\gamma\gamma}\sim 4\times 10^{-4}$~GeV$^{-1}$ with $10^{12}$ MOT using the visible channel, and $10^{14}$ MOT will further close the gap between beam dump and collider experiments such as BESIII. Even with the 2023 run, new parameter space will start to be excluded between NA64e and OPAL. The invisible channel will reach similar sensitivity at lower masses, which are partly ruled out by existing experiments. However, complementarity between visible and invisible channels could still be achieved for $10^{14}$ MOT.  

The 2023 run also allows to exclude the $g_{a\mu\mu}\gtrsim 4\times 10^{-3}$ GeV$^{-1}$ for $m_a\lesssim 0.2$~GeV and $10^{12}$ ($10^{14}$) MOT could exclude $g_{a\mu\mu}\gtrsim 5.7\times 10^{-4}$~$(5.7\times 10^{-5})$~GeV$^{-1}$. For a flavor-dependent dark sector, the sensitivity on $(g_{Z'}g_\chi)^2$ simple scales as $N_{\rm MOT}^{-1}$. The 2023 run ($10^{12}$ MOT) will lead to about one (two) order of magnitude enhancement in the sensitivity.

Apart from $L_\mu-L_\tau$, a muon beam dump also paves the way to the direct probe of the flavor structure of BSM physics, particularly involving the second and third generation leptons. For example, in the case of the flavor-violating interactions, a muon-to-tau transition will be followed by the subsequent tau decay in ECAL. If tau decays leptonically to a muon that passes the muon trackers, it will satisfy the trigger in NA64$\mu$. This represents one of the few ways to explore flavor transition involving a tau lepton. Axion with tau coupling in addition to muon will also produce a pair of taus in the final states and be identified exclusively at NA64$\mu$. With the advent of muon beam dumps and the upcoming muon colliders, we are not just entering an era of precise study, but also opening the window to the possibilities of flavor-specific interactions. 

\section*{Acknowledgement}
We thank Paolo Crivelli and Laura Molina Bueno for useful correspondence on the NA64$\mu$ experiment. We also acknowledge Ui Min for useful correspondence.
HL and ZL 
are supported by the 
National Natural Science Foundation of China (NSFC) under Grant Nos.\ 12275128 
and 12147103. 
NS is supported by the NSFC Project Nos. 12475110, 12347105, 12441504 and 12447101.

\appendix
\section{Decay probability of long-lived axion}\label{sec:appendix-decay}

The ECAL consists of 150 layers of lead and plastic scintillator (Sc) plates \cite{Andreev:2024yft}. Each layer contains 1.5 mm thick lead and 1.5 mm thick Sc, which combine into 40$X_0$ of lead with $X_0=0.56$~cm. Among them, 4$X_0$ are pre-shower detector, resulting in a net target length of $L_T=20.16$~cm for muon scattering~\cite{Sieber:2024vhx}. We assume that the probability of scattering in each layer of lead is the same. This is reasonable given the small energy loss of muon in ECAL (about 0.5~GeV)~\cite{Groom:2001kq}.

We label the layers in the ECAL from the last layer to the front as the 
$0$-th to the $135$-th layer (excluding the pre-shower layers).
If an axion is produced in the last layer of the ECAL 
(the $0$-th layer),  
the probability for the axion to decay before the end of the last HCAL (i.e. visible decay) is
\begin{equation}
    P_{0, {\rm v}}=1-e^{-L_{\rm E-H}/l_a}\,,
\end{equation}
where 
$l_a$ is the axion decay length, and  
we have ignored the uncertainty for the location of the muon scattering inside the last layer. $L_{\rm E-H}=10.65~\rm{m}$ is the distance from the end of the ECAL to the end of the last HCAL~\cite{Andreev:2024yft}. If the axion 
that is produced in the $0$-th layer of the ECAL 
escapes the last HCAL without decaying (i.e. invisible decay), the probability gives 
\begin{equation}
    P_{0, {\rm inv}}=e^{-L_{\rm E-H}/l_a}\,.
\end{equation}
We denote by $P_{k, {\rm v(inv)}}$ the probability that an axion produced in the $k$-th layer of the ECAL 
remains invisible while traversing the ECAL 
and subsequently produces a visible (invisible) signature between the last layer of the ECAL and the end of the last HCAL.
Thus, one has 
\begin{equation}
P_{k, {\rm v(inv)}}=e^{-kd_0/l_a} P_{0, {\rm v(inv)}},     
\end{equation}
where  $d_0=3~\mathrm{mm}$ is the thickness of one lead-Sc layer.
With the assumption that the probability of scattering in each layer
is the same, we obtain the average visible (invisible) decay probability via 
$
{P}_{\mathrm{v(inv)}}=\frac{1}{136}\sum_{k=0}^{135}P_{k,\rm v(inv)}\
$.

\section{Virtual photon flux}

The virtual photon flux in Eq. (4) in the main text can be integrated out as
\begin{equation}
    \zeta=-Z^2\frac{b^4d^2}{(-1+b^2d)^3}(\zeta_1+\zeta_2)\,,
\end{equation}
where
\begin{equation}
\begin{split}
   &\zeta_1= \frac{(-1+b^2d)(1+b^2(d+2t_{\mathrm{max}}))(t_{\mathrm{max}}-t_{\mathrm{min}})}{(d+t_{\mathrm{max}})(1+b^2t_{\mathrm{max}})}\,,\\
   &\zeta_2=(1+b^2(d+2t_{\mathrm{min}}))\ln\left[\frac{(d+t_{\mathrm{max}})(1+b^2t_{\mathrm{min}})}{(d+t_{\mathrm{min}})(1+b^2t_{\mathrm{max}})}\right]\,.
\end{split}
\end{equation}

\bibliography{NA64mu}

\end{document}